\documentclass{article}
\usepackage{PRIMEarxiv}

\usepackage[utf8]{inputenc} 
\usepackage[T1]{fontenc}    
\usepackage{hyperref}       
\usepackage{url}            
\usepackage{booktabs}       
\usepackage{amsfonts}       
\usepackage{nicefrac}       
\usepackage{microtype}      
\usepackage{lipsum}
\usepackage{fancyhdr}       
\usepackage{graphicx}       
\graphicspath{{media/}}     
\usepackage{eucal}
\usepackage{longtable}
\usepackage{multirow}

\newcommand*{\LargerCdot}{\raisebox{-0.6ex}{\scalebox{2}{$\cdot$}}}

\title{Graph Auto-Encoders for Financial Clustering}
\author{{\hspace{1mm}Edward Turner}\thanks{Research Intern at the Oxford-Man Institute of Quantitative Finance, University of Oxford. Supervised by Prof. Mihai Cucuringu (University of Oxford, Department of Statistics, OxCSML) and Prof. Xiaowen Dong (University of Oxford, Department of Engineering Science, MLRG).} \\
	University of Oxford\\
	\href{mailto:edward.turner@mansfield.ox.ac.uk}{\texttt{edward.turner@mansfield.ox.ac.uk}}\\
	\href{mailto:edward.turner01@outlook.com}{\texttt{edward.turner01@outlook.com}}\\
}

\begin{document}
\maketitle
\begin{abstract}
Deep learning has shown remarkable results on Euclidean data (e.g. audio, images, text) however this type of data is limited in the amount of relational information it can hold. In mathematics we can model more general relational data in a graph structure while retaining Euclidean data as associated node or edge features. Due to the ubiquity of graph data, and its ability to hold multiple dimensions of information, graph deep learning has become a fast emerging field. We look at applying and optimising graph deep learning on a finance graph to produce more informed clusters of companies. Having clusters produced from multiple streams of data can be highly useful in quantitative finance; not only does it allow clusters to be tailored to the specific task but the culmination of multiple streams allows for cross source pattern recognition that would have otherwise gone unnoticed. This can provide financial institutions with an edge over competitors which is crucial in the heavily optimised world of trading. In this paper we use news co-occurrence and stock price for our data combination. We optimise our model to achieve an average testing precision of $78\%$ and find a clear improvement in clustering capabilities when dual data sources are used; cluster purity rises from $32\%$ for just vertex data and $42\%$ for just edge data to $64\%$ when both are used in comparisons to ground-truth Bloomberg clusters. The framework we provide utilises unsupervised learning which we view as key for future work due to the volume of unlabelled data in financial markets.
\end{abstract}

\section{Introduction}
\subsection{Background}
In quantitative finance research, market and trading patterns are identified, models are built on those patterns, and the information is used to predict future price and direction of securities. To identify these patterns is no easy task; funds have spent billions of dollars developing models all with the goal of predicting market performance, none perfect. The reason this task is so difficult is because of inherently unstable factors and a complicated market outlook. The stock market reflects many influences, ranging from news and current affairs all the way to weather. Having more informed and tailored clusters of companies can allow models to be more specific, gain greater insight and produce better results.

Recent rapid development in technology and automation of processes has led to the use of machine learning in many fields, including finance. Researchers have proven that machine learning methods are able to analyse a large amount of data in a short period of time with greater accuracy and effectiveness \cite{app9245574}. Due to noise and dependencies within most data, which is especially prevalent in the stock market, identifying significant information is a very difficult task but it can be handled by machine learning models \cite{app9245574}.

\subsection{Problem Setup}
Our first aim is to build a finance graph from given data sources and then clean and optimise the data to increase model prediction ability. Next we aim to train a graph auto-encoder on our finance graph with the purpose of using the trained encoder to produce a latent representation of the finance graph. Then from the latent representation we intend to use k-means clustering to produce highly informed clusters of the graph vertices. Overall, our goal is to show that graph auto-encoders are capable of capturing both edge data and vertex information to produce more informed clusters of companies based on finance data. Highly informed clustering like this can be very helpful for downstream tasks, such as a fund manager wanting to minimise risk. To do this they could take clusters customised to their funds needs and diversify the funds portfolio across the clusters, spreading risk.

\subsection{Summary of Main Contributions}
In this paper we investigate how graph auto-encoders can produce more informed clusters of companies; to the best of our knowledge this is the first time graph auto-encoders have been applied in a financial context. We provide data cleaning and optimisation methods for such a task and then an overview on applying and training a graph auto-encoder. We also provide code for k-fold cross validation in PyTorch Geometric which we believe to be the only publicly available. We then investigate how our model compares to classical graph clustering methods. Finally, we compare our model to the graph auto-encoder trained on just one data source to highlight the additional performance in leveraging information from multiple sources simultaneously.

\section{Related literature}
\subsection{Graph Representation Learning}
The goal of graph representation learning is to take a feature rich graph (e.g. a social network with edges representing friendships and vertices holding information on each individual) and then in a low-dimensional vector space, the \textbf{latent space}, produce a feature representation embedding of the graph, the \textbf{latent representation}. This provides a better environment for inference of the data where commonly one would be tasked with \textbf{vertex classification}, the prediction of certain vertices labels for some or all of the vertices, or \textbf{relation prediction}, the prediction of whether two nodes are linked by an edge or not \cite{grl}. In this paper we train our model via relation prediction.

\subsection{Graph Neural Networks}
With the recent success and advances of machine learning, and specifically deep learning, cutting edge graph representation learning mostly falls into the category of \textbf{graph neural networks} (GNNs). Here the challenge of applying deep learning to graph data (i.e. non-Euclidean data) is overcome\footnote[1]{One may wonder why we can not simply feed the graph adjacency matrix into a classic neural network. The issue with this approach is it depends on the arbitrary labelling of the nodes and is thus not permutation invariant \cite{grl}.}. GNNs work via a message passing framework with multiple layers, similar to that of a classic neural network. The layers of a GNN all contain $N$ nodes and may be represented as matrices, $\mathbf{H}^{(0)}\in\mathbb{R}^{N\times{D_0}}$, ..., $\mathbf{H}^{(L)}\in\mathbb{R}^{N\times{D_L}}$. We use $\mathbf{h}^{k}_i$, the i\textsuperscript{th} row of $\mathbf{H}^{(k)}$, to represent the i\textsuperscript{th} node in the k\textsuperscript{th} layer which corresponds to the i\textsuperscript{th} vertex of $\CMcal{G}$.

A GNN is initialised by the following iteration:

\begin{equation}
    \mathbf{H}^{(0)} = \mathbf{X}\, ,
\end{equation}
\begin{equation}
\label{gnnequation}
    \mathbf{h}^{(k+1)}_i = \mathrm{Update}^{(k)}(\mathbf{h}^{(k)}_i, \mathrm{Aggregate}^{(k)}(\{\mathbf{h}^{(k)}_j : e_{i, j} \in \CMcal{E}\}))\, ,
\end{equation}

where the $\mathrm{Updates}$ and $\mathrm{Aggregates}$ are chosen differentiable functions \cite{grl}. The first layer is set to be the vertex features and then for each following layer a vertex's embedding is based on its previous embedding updated with an aggregation of its neighbouring vertices previous embeddings. We then take $\mathbf{H}^{(L)}$ to be our final low-dimension embedding of the graph (the latent representation). Once a loss function is added the model can be trained via back-propagation and stochastic gradient descent by tuning the weights associated to the $\mathrm{Updates}$ and $\mathrm{Aggregates}$ functions.

The models we use in this paper add a slight simplification by joining each $\mathrm{Update}^{(k)}$ and $\mathrm{Aggregate}^{(k)}$ pair into one function. This is known as a \textbf{self-loop} approach and often alleviates overfitting \cite{grl}. In a self-loop model Equation~\hyperref[gnnequation]{2} becomes the following:

\begin{equation}
\label{selfloopequation}
    \mathbf{h}^{(k+1)}_i = \mathrm{Aggregate}^{(k)}({\{\mathbf{h}^{(k)}_i}\} \cup \{\mathbf{h}^{(k)}_j : e_{i, j} \in \CMcal{E}\})\, ,
\end{equation}

thus now a vertices' current embedding is viewed with equal importance as its neighbour's embeddings when updating for the next layer.

\subsection{Graph Convolutional Networks}
\textbf{Convolutional neural networks} (CNNs) have seen tremendous success in practical applications \cite{Goodfellow-et-al-2016}. It was only a matter of time before a similar convolution methodology was brought to GNNs and in $2016$ Kipf and Welling released two papers providing a new framework, the \textbf{graph convolutional network} (GCN) \cite{DBLP:journals/corr/KipfW16} \cite{kipf2016variational}.

The GCN architecture uses a layer-wise propagation rule, motivated by a first-order approximation of spectral graph convolutions. The model is able to encode both the local graph structure and the vertex features. In a number of experiments on citation networks and a knowledge dataset, it was shown that the model outperforms recent related methods such as DeepWalk and Planetoid \cite{DBLP:journals/corr/KipfW16}.

The GCN architecture has also been shown to achieve competitive results for unsupervised learning on graph-structured data. The \textbf{graph auto-encoder} (GAE) model is similar to a normal auto-encoder but with an inference model (encoder) based on a multi-layer GCN. This model was able to learn meaningful latent representations during a relation prediction task on three popular citation network datasets. It also gives better predictive performance than spectral clustering and DeepWalk \cite{DBLP:journals/corr/KipfW16}. In contrast to related methods, GAEs can also incorporate vertex features which results in a performance boost.

\subsection{Machine Learning on Finance Graphs}
With the development of neural network models, deep learning has become a popular way to solve the stock prediction problem. Despite the exciting advances in machine learning applied to finance, many current approaches are limited to using technical analysis to capture historical trends of each stock price and thus limited to certain experimental setups to obtain good prediction results. By adapting to a graph structure and incorporating company knowledge graphs directly into their predictive model, Matsunaga et \nonfrenchspacing{al. were} able to obtain significant increases to return and Sharpe ratio when compared to market benchmarks \cite{DBLP:journals/corr/abs-1909-10660}.

By similarly capitalising on multiple streams of data, Wan et al. were able to extract subtleties of market information that may have otherwise gone unnoticed. They investigated how combining news co-occurrence with quarterly company sentiment data can provide insight into market returns \cite{DBLP:journals/corr/abs-2011-06430}. They found that there exists a weak but statistically significant association between strong media sentiment and abnormal returns and volatility. Such an association is more significant at the level of individual companies, but nevertheless remains visible at the level of sectors or groups of companies.

\section{Methodology}
\subsection{Preliminary Definitions and Notation}
A \textbf{graph} $\CMcal{G}$ is a pair: $\CMcal{G} = (\CMcal{V},\CMcal{E})$, comprising of a set of \textbf{vertices}, $\CMcal{V} = \{v_1, ..., v_{N}\}$, and a set of \textbf{edges}, $\CMcal{E} = \{e_{i_1, j_1}, ..., e_{i_{M}, j_{M}}\}$. Each edge, $e_{i, j}$, connects the pair $(v_i, v_j)$ where $v_i, v_j \in \CMcal{V}$. $\CMcal{G}$ may be fully expressed by its \textbf{adjacency matrix}, $\mathbf{A}=(a_{i,j})\in\mathbb{R}^{N\times{N}}$, where $N = |\CMcal{V}|$ and $a_{i,j}=1$ if  $e_{i, j}\in\CMcal{E}$ or $0$ otherwise.

A graph is \textbf{weighted} if each edge, $e_{i, j}$, has an associated weight, $w_{i, j} \in \mathbb{R}$. In this case the adjacency matrix has entries $a_{i,j}=w_{i,j}$ if  $e_{i, j}\in\CMcal{E}$ and $0$ otherwise.

A graph is \textbf{featured} if it has an associated \textbf{feature matrix}, $\mathbf{X}=(x_{i,j})\in\mathbb{R}^{N\times{D}}$, where $D =$ the dimension of the feature space. The i\textsuperscript{th} row of $\mathbf{X}$ corresponds to the D features associated to the i\textsuperscript{th} vertex of $\CMcal{G}$.

Given two vectors, $\mathbf{v}_1, \mathbf{v}_2\in\mathbb{R}\textsuperscript{n}$, we define their \textbf{cosine distance} ($\mathrm{CD}$) as:

\begin{equation}
    \mathrm{CD}(\mathbf{v}_1, \mathbf{v}_2) = \frac{\mathbf{v}_1 \LargerCdot \mathbf{v}_2}{||\mathbf{v}_1||\,||\mathbf{v}_2||}\, ,
\end{equation}

where \small $||\cdot ||$ \normalsize is the Euclidean norm. This metric corresponds to a measure in similarity of the vectors headings in $n$ dimensional space.

\subsection{Symmetric Matrix Normalisation}
For an adjacency matrix $\mathbf{A}=(a_{i,j})\in\mathbb{R}^{N\times{N}}$ we define $\mathbf{A}^{*}$ to be the \textbf{symmetric normalisation} of $\mathbf{A}$. That is:

\begin{equation}
    \mathbf{A}^{*} = \mathbf{D}^{-\frac{1}{2}}\mathbf{A}\mathbf{D}^{-\frac{1}{2}}\, ,
\end{equation}

where $\mathbf{D}=({d}_{i,j})\in\mathbb{R}^{N\times{N}}$ is the weighted vertex degree matrix. Explicitly, $\mathbf{D}$ is diagonal with ${d}_{ii} = \sum_j \mathbf{A}_{ij}$. Thus we have:

\begin{equation}
    a_{i,j}^{*} = \frac{a_{i,j}}{\sqrt{d_{i,i}}\sqrt{d_{j,j}}}\, .
\end{equation}

We may interpret the graph represented by $\mathbf{A}^{*}$ as normalising each edge of $\mathbf{A}$'s graph by the geometric mean of its start and end vertices weighted degrees. This is opposed to the simpler one sided normalisation's $\mathbf{D}^{-1}\mathbf{M}$ or $\mathbf{M}\mathbf{D}^{-1}$ which normalises on just the start or end vertex weighted degree respectively.

\subsection{Graph Convolutional Network Encoder}
For our Graph Auto-Encoder model we use the following GCN architecture with two hidden layers as our encoder\footnote[2]{See Section~\hyperref[sec:K-fold_Cross_Validation]{5.2} for why two hidden layers are chosen.}. We take inspiration from Kipf and Welling \cite{DBLP:journals/corr/KipfW16} with the following layer-wise iteration for our GCN:

\begin{equation}
    \mathbf{H}^{(0)} = \mathbf{X}\, ,
\end{equation}
\begin{equation}
  \mathbf{H}^{(k+1)} = \sigma\!\left(\mathbf{\tilde{A}}^{*}\mathbf{H}^{(k)}\mathbf{W}^{(k)}\right)\, .
\end{equation}

Here, $\mathbf{W}^{(k)}$ is the trainable weight matrix for the k\textsuperscript{th} layer of the GCN. It is due to the sharing of a weight matrix per layer that the "convolutional" name is justified. $\sigma(\cdot)$ is a coordinate-wise non-linear activation function, which we take to be the $\mathrm{ReLU}$ function: $\mathrm{ReLU}(\cdot) = \max(0,\cdot)$.

$\mathbf{\tilde{A}} = \mathbf{A} + \mathbf{I}_N$ is the adjacency matrix of $\CMcal{G}$ with added self-loops, which we then symmetrically normalise, producing $\mathbf{\tilde{A}}^{*}$. Adding self-loops allows us to apply the self-loop method in Equation~\hyperref[selfloopequation]{3} to counteract overfitting.  We normalise $\mathbf{\tilde{A}}$ as otherwise multiplying by it will change the scale of the feature vectors. The reason we choose symmetric normalisation for our normalisation method is that the dynamics are more interesting as it no longer amounts to mere averaging of neighbouring nodes \cite{DBLP:journals/corr/KipfW16}. Furthermore, symmetric normalisation has proven to be one of the most effective baseline GCN architectures \cite{grl}.

\subsection{Inner Product Decoder}
Due to our data being unlabelled we use an auto-encoder and thus also need a decoder (alongside the GCN which will act as an encoder). We use a pairwise decoder, $\mathrm{DEC}$, where:

\begin{equation}
  \mathrm{DEC}(\mathbf{h}^{(L)}_i, \mathbf{h}^{(L)}_j) =  \sigma\!\left(\mathbf{h}^{(L)}_i\LargerCdot \mathbf{h}^{(L)}_j\right)\, .
\end{equation}

Here $\sigma(\cdot)$ is again a non-linear activation function, but this time we take it to be the sigmoid function: $\mathrm{S}(x) = \frac{1}{1+e^{-x}}$.

\subsection{Graph Auto-Encoder}
 Combining our GCN encoder with the inner product decoder we have our final GAE model. From our model we produce a reconstruction matrix $\mathbf{S}=(s_{i,j})\in\mathbb{R}^{N\times{N}}$ where $s_{i,j} = \mathrm{DEC}(\mathbf{h}^{(L)}_i, \mathbf{h}^{(L)}_j)$. Our reconstruction loss function then compares $\mathbf{A}$, the original adjacency matrix, to $\mathbf{S}$ by computing their binary cross entropy \cite{DBLP:journals/corr/KipfW16}. We then use Adam as the optimiser to train our model as this is an industry standard for stochastic gradient descent \cite{kingma2017adam}.

\section{Data Description and Optimisation}
\subsection{News Co-occurrence Data}
The first of the two datasets we use is the news co-occurrence data provided to us by Xingchen Wan \cite{DBLP:journals/corr/abs-2011-06430}. Wan et al. selected $87$ companies and through \textbf{natural language processing} (NLP) techniques the companies were detected from financial news articles on Reuters across $2007$ with entities that co-reference the same company, i.e. \textit{AAPL} and \textit{Apple}, merged as one. From this a news coverage matrix, $\mathbf{M}=(m_{i,j})\in\mathbb{R}^{87\times{12,311}}$, was constructed where each row represents a company and each column represents one of the $12,311$ news articles from $2007$. $m_{i,j} = 1$ if company i appears in news j and $0$ otherwise\footnote[3]{This data was pre-processed thanks to Wan et al. \cite{DBLP:journals/corr/abs-2011-06430}.}.

\subsection{Market Close Data}
For our second dataset we use the $2007$ daily market close data for a plethora of listed companies. The bulk of the market close data comes from CRSP (for companies on the NYSE) and Wharton (for companies not on NYSE but on the S\&P $1500$)\footnote[4]{This data was provided thanks to Prof. Cucuringu.}. Any of the $87$ companies that weren't covered by the CRSP or Wharton datasets but had their $2007$ market close data on Yahoo! finance we took from there instead. Unfortunately not all of Wan et al.'s $87$ companies were within out market close data so we had to take the intersection, resulting in $72$ companies. 

We collated the data from the multiple sources into one file. This had the $72$ companies with daily market close price for each company's stock across the $251$ days in the $2007$ financial year. After we collated the data we found $3$ days of feature data for one node were missing (due to a storage error on Yahoo! finance). To account for the missing days we took them to be the arithmetic meaning of their neighbouring days.

\subsection{Finance Graph}
\begin{figure}[t]
\centering
\includegraphics[scale=0.9]{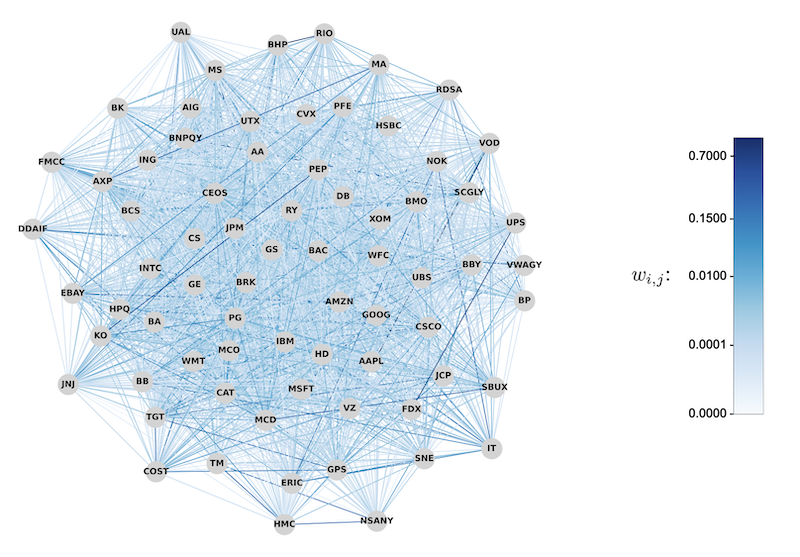}
\caption{\label{fig:NetworkX_graph}News co-occurrence graph, containing the 72 companies used with cosine distance of the news co-occurrence for edges. (To map the edge weight colours to a more visually pleasing scale the mapping $f(w_{i, j})=w_{i, j}^{1/7}$ was used.)}
\end{figure}

\subsubsection{Edges and Vertices}
The news co-occurrence dataset captures pairwise relations between companies and thus we use it to construct our graph edges and their associated weights. First we construct $\mathbf{\tilde{M}}\in\mathbb{R}^{72\times{12,311}}$, which is just $\mathbf{M}$ after deleting the rows of the $15$ companies we are no longer using.

From $\mathbf{\tilde{M}}$ our news co-occurrence graph was constructed. The vertices of the graph are the $72$ companies used and the edge weights are defined by the cosine distance of the associated rows from $\mathbf{\tilde{M}}$ of the vertices each edge connects. That is the edge weight, $w_{i, j}$, between a pair of vertices i and j is defined by:

\begin{equation}
    w_{i,j} = \mathrm{CD}(\mathbf{\tilde{m}}_i, \mathbf{\tilde{m}}_j)\, ,
\end{equation}

where $\mathbf{\tilde{m}}_i$ is the i\textsuperscript{th} row of $\mathbf{\tilde{M}}$. If $w_{i,j} = 0$ then there is no edge between vertex i and j.  Figure~\ref{fig:NetworkX_graph} shows the news co-occurrence graph plotted in NetworkX. One can note expected relationships such as Pepsi (PEP) and Coca-Cola (KO) or Starbucks (SBUX) and McDonald's (MCD) can be seen, showing the graph has indeed captured real relational information from the Reuter's news articles.

Let $\mathbf{A}_{nc}$ denote the adjacency matrix for our news co-occurrence graph. Before we set $\mathbf{A}_\CMcal{F}$, the adjacency matrix of our finance graph $\CMcal{G}_\CMcal{F}$, equal to $\mathbf{A}_{nc}$ we carry out two data optimisation steps. Our first step is removing entries of $\mathbf{A}_{nc}$ to produce $\mathbf{A}_{nc}'$, a new adjacency matrix whose associated graph has half of all possible edges. We do this by calculating the median of all entries in $\mathbf{A}_{nc}$ and then setting any entry with a value less than this to zero (i.e. removing the edges of least weight until we have a graph with half of all possible edges). The reason for doing this is in the training of our GAE model we assign negative weights to any edge that does not exist and balancing the number of existing edges with not existing edges gives the best performance. $\mathbf{A}_{nc}$ has $1986/2556$ edges and $\mathbf{A}_{nc}'$ has, by design, $1278/2556$ edges.

For our second step we scale all entries of $\mathbf{A}_{nc}'$, producing $\mathbf{A}_{nc}''$ an adjacency matrix whose entries have a mean value of $1$. We do this by simply taking the mean of the entries of $\mathbf{A}_{nc}'$ and then dividing each entry by the mean. The reason we do this is the mean edge weight of $\mathbf{A}_{nc}'$ is $0.00873$. This is an issue as in the setup of our GCN when $\mathbf{\tilde{A}}$ is produced the addition of self-loops would dominate almost all the message passing (since the self-loops edges have a weight of $1$ while the average weight of all other edges is $0.00873$), effectively rendering the edges of our graph useless.

\begin{figure}[t]
\centering
\includegraphics[scale=0.6]{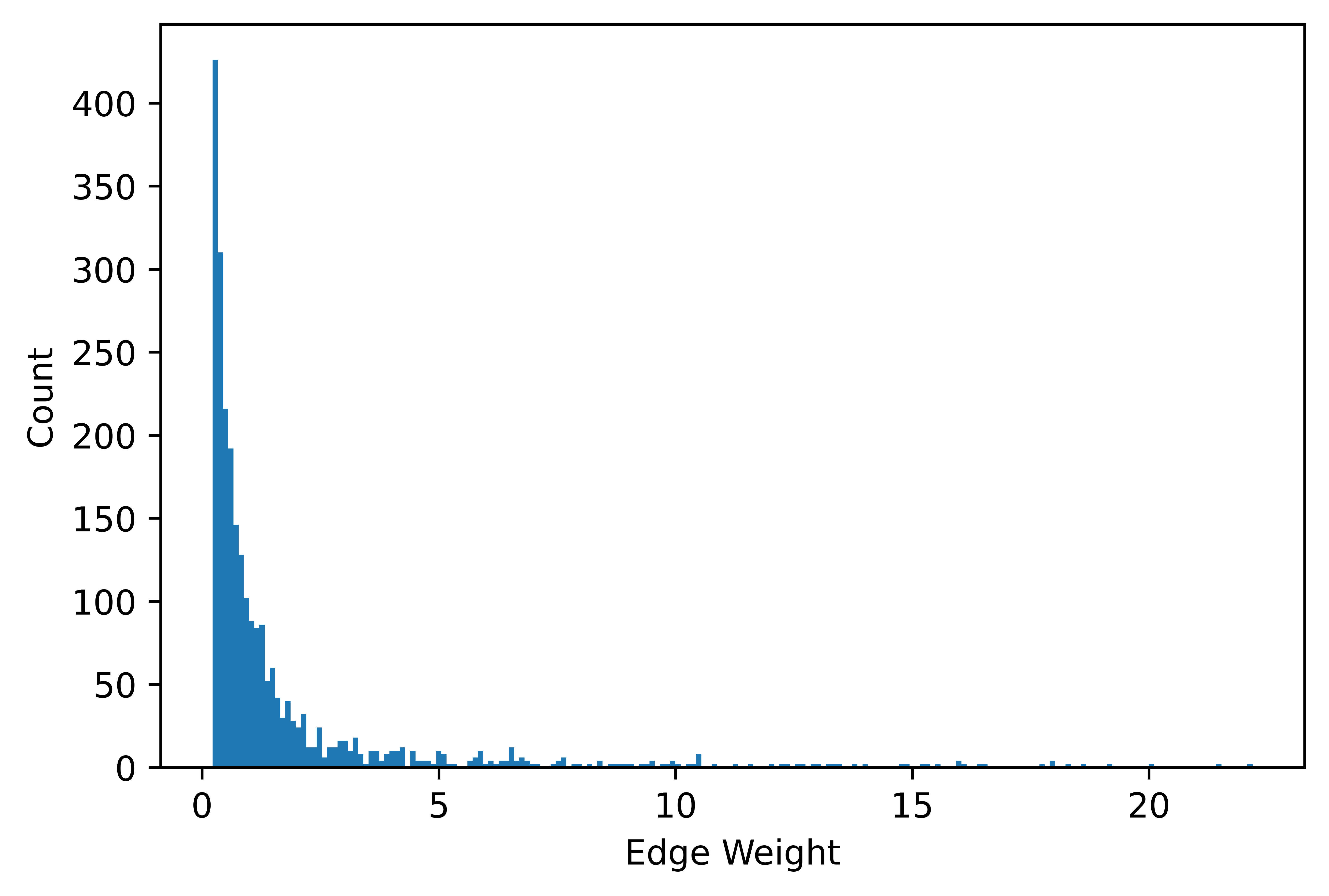}
\caption{\label{fig:edge_weight_histogram}Histogram with $200$ bins showing the distribution of the edge weights of $\CMcal{G}_\CMcal{F}$. Anomalies above $30$ were removed for better plotting (in total this removed $16/5184$ with the largest anomaly being an edge weight of $71.5$).}
\end{figure}

Now that edges below the median weight have been removed and the remaining edges have been scaled by the mean weight we may take $\mathbf{A}_\CMcal{F}\in\mathbb{R}^{72\times{72}}$, the adjacency matrix of our finance graph $\CMcal{G}_\CMcal{F}$, to be equal to $\mathbf{A}_{nc}''$. Figure~\ref{fig:edge_weight_histogram} shows a histogram of the entries of $\mathbf{A}_\CMcal{F}$ (i.e. the distribution of the edge weights in our finance graph).

\subsubsection{Vertex Features}
Once we collated our market close data we chose to use the \textbf{previous Close Close} ($\mathrm{pvCLCL}$) metric, that is the overnight linear return from yesterdays (i.e. previous) close (adjusted for splits and dividends) until today’s close.

Explicitly:

\begin{equation}
    \mathrm{pvCLCL}(v_{i,j}) = \frac{c_{i,j}-c_{i,j-1}}{c_{i,j-1}} \, ,
\end{equation}

where $v_{i,j}$ is stock i on day j and $c_{i,j}$ is the respective stock's close price. We find this percentage change metric to be a much better scale to compare between stocks rather than just stock close price.

From our $\mathrm{pvCLCL}$ data we construct a feature matrix, $\mathbf{X}=(x_{i,j})\in\mathbb{R}^{72\times{250}}$, where $x_{i,j} = \mathrm{pvCLCL}(v_{i,j})$. Here the feature matrix has a $250$ dimensional feature space due to there being $250$ $\mathrm{pvCLCL}$ instances from the $251$ market close prices in the $2007$ financial year.

After plotting a heatmap for $\mathbf{X}$ (see Figure~\ref{fig:original_heatmap} in the Appendix) we see the CeCors (CEOS) feature data to be a clear anomaly. Upon inspection this is indeed the case with CeCors having daily swings up to $183\%$ (Oct. 4\textsuperscript{th}). After checking the data had been correctly inputted we concluded hard clipping windorisation (setting a max and min possible value for each $x_{i,j}$) of the data was the most appropriate next step. We used a $[-0.1, 0.1]$ window, capping daily $\mathrm{pvCLCL}$ swings to no more than $10\%$. $99.52\%$ ($17,913/18,000$) of the vertex feature data was already within this window thus we are only capping true anomalies.

\begin{figure}[t]
\centering
\includegraphics[scale=4.2]{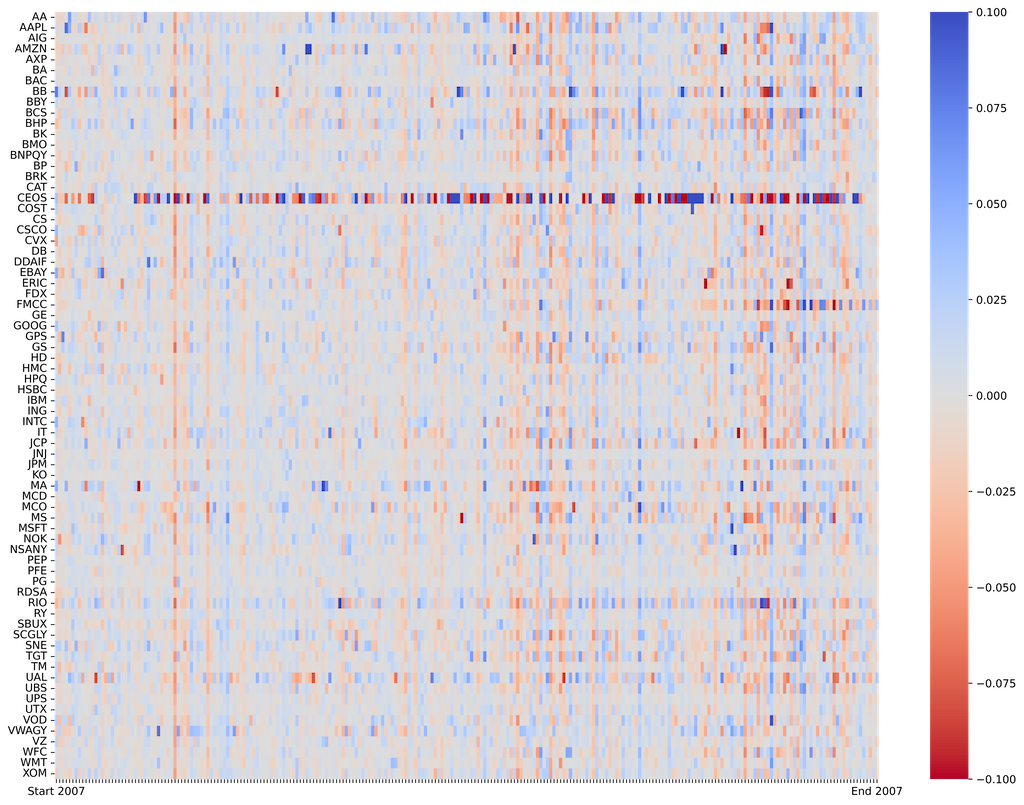}
\caption{\label{fig:windorised_heatmap}Heatmap of the windorised $\mathrm{pvCLCL}$ vertex features.}
\end{figure}

We denote the windorised node features as $\mathbf{X}'=(x_{i,j}')\in\mathbb{R}^{72\times{250}}$, where $x_{i,j}' = \mathrm{sign}(x_{i,j}) \mathrm{min}(|x_{i,j}|, 0.1)$. Figure~\ref{fig:windorised_heatmap} shows a heatmap for our windorised data. The clearer blue/red vertical strips show the market as a whole increasing/decreasing while the horizontal strips show the general change for a specific company. We note that CeCors still exhibits the strongest colours due to its high volatility.

While windorised $\mathrm{pvCLCL}$ proved to be better than close price as an inter stock comparison metric, we found the daily discrepancies were still too large. To improve results we normalised our vertex $\mathrm{pvCLCL}$ data across each day, producing a normalised vertex feature matrix $\mathbf{X}''=(x_{i,j}'')\in\mathbb{R}^{72\times{250}}$. The entries of $\mathbf{X''}$ were calculated in the following way:

\begin{equation}
    x_{i,j}'' = \frac{x'_{i,j} - \mu_j}{\sigma_j}\, ,
\end{equation}

where $\mu_j$ and $\sigma_j$ are the mean and standard deviation of the j\textsuperscript{th} column of $\mathbf{X'}$ respectively.

\begin{figure}[t]
\centering
\includegraphics[scale=0.6]{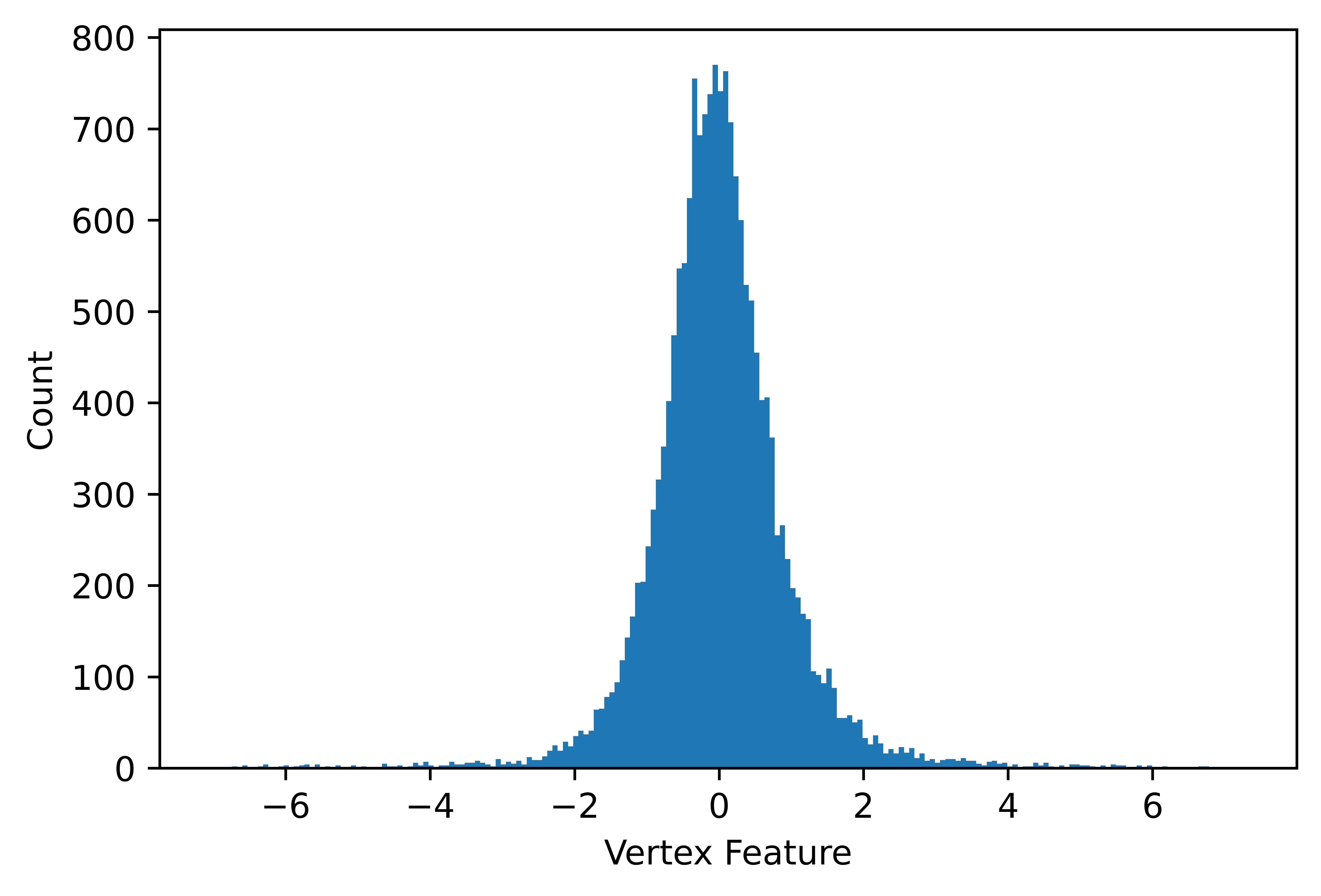}
\caption{\label{fig:vertex_feature_histogram}Histogram with $200$ bins showing the distribution of the vertex features of $\CMcal{G}_\CMcal{F}$.}
\end{figure}

Having normalised the data we plotted it as a histogram as shown in Figure~\ref{fig:vertex_feature_histogram}. Since the vertex feature data had now been optimised we set $\mathbf{X_\CMcal{F}}$, the vertex feature matrix of our finance graph, to be equal to $\mathbf{X}''$. We now have our featured and weighted finance graph $\CMcal{G}_\CMcal{F}$.

\section{Numerical Experiments and Results} 
\subsection{Model Regularisation}
\begin{figure}[t]
\centering
\includegraphics[scale=0.48]{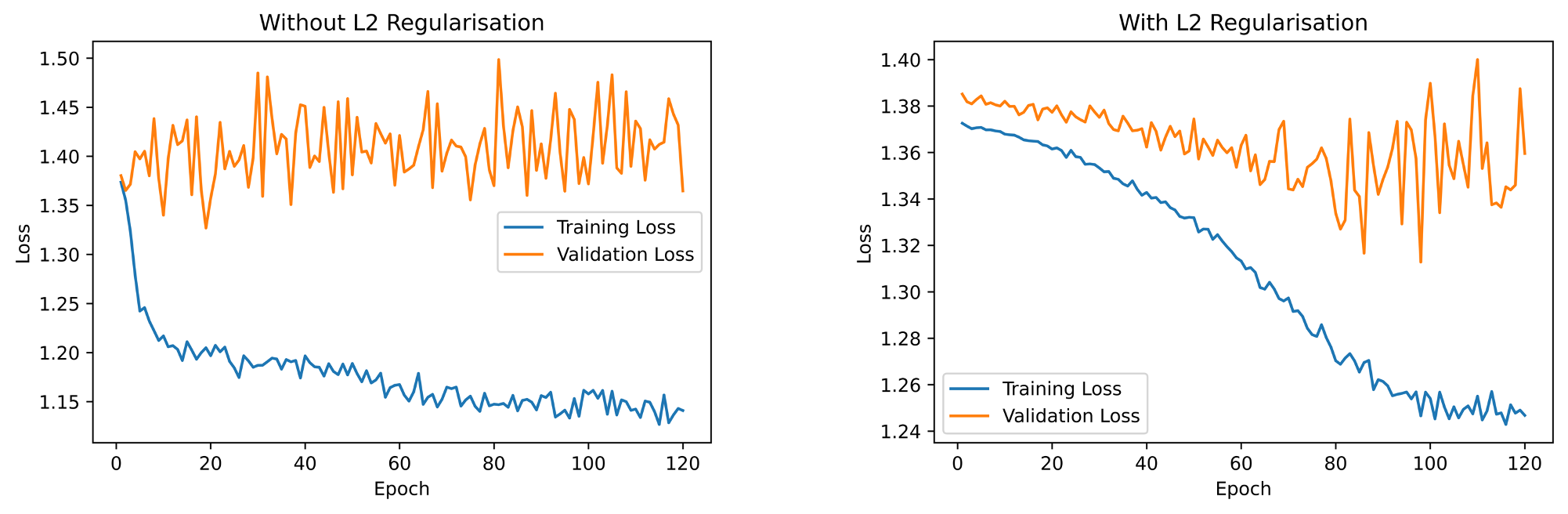}
\caption{\label{fig:train_vs_val_loss_plots}Plots of training our model without and with L2 regularisation. Without L2 regularisation shows clear overfitting as the validation loss does not decrease on average. Once L2 regularisation is added we see a clear decrease in the validation as well as the training loss.}
\end{figure}

When implementing our GAE model on the finance graph we use $20\%$ of our edge data for testing and $16\%$ for validation. When first running the model without any regularisation methods we noticed that despite using the self-loop method the model was drastically overfitting and just learning noise in the training data. Initially to overcome this we added L2 (ridge) regularisation with a weight decay of $0.01$ to our model \cite{DBLP:journals/corr/abs-1205-2653}. This adds a penalty to the weights of our model; forcing them to remain relatively small and thus reducing overfitting. Figure~\ref{fig:train_vs_val_loss_plots} shows how much of a difference L2 regularisation makes to our model on a training run of 120 epochs.

Figure~\ref{fig:train_vs_val_loss_plots} also shows how our model experiences high variation in the validation loss once training loss reaches a certain threshold. To overcome this we designed a custom early stopping method for the number of epochs. Rather than Kipf and Welling's early stoping method (they stop their model if validation loss fails to decrease for 10 epochs \cite{DBLP:journals/corr/KipfW16}) we implement a method that takes a rolling average to alleviate the variation seen. Specifically we wait until a threshold number of epochs (which we take to be 80) and then look at the arithmetic mean of the n previous validation losses and compare this to the mean of the n before them. If the mean has failed to decrease we stop training\footnote[5]{We end up taking n $= 30$,  this choice is discussed in Section~\hyperref[sec:K-fold_Cross_Validation]{5.2}.}.

\subsection{K-fold Cross Validation}
\label{sec:K-fold_Cross_Validation}
Our model has a variety of hyperparameters, namely: the number of hidden layers, the dimension of the hidden layers, the dimension of the output (latent space), the L2 regularisation weight decay rate, the learning rate and the value for n in our early stopping method. Given the amount of hyperparameters and relatively small amount of data we viewed k-fold cross validation as the best optimisation method for our model.

We use 5-fold cross validation. When coding this care was taken to ensure the same permutation of the edges was used in creating the test, train, validation split so that we could not inadvertently be using a test edge which had previously been in a train or validation set.

We considered $1$ or $2$ hidden layers with a hidden dimension of $64$ or $128$ for the $1$ layer model and hidden dimensions of $(64, 16)$, $(64, 32)$, $(64,64)$, $(128, 16)$, $(128,32)$ or $(128,64)$ for the $2$ layer model. Alongside these we tested an output dimension of $8$, $16$ or $32$. We also considered $0.1$, $0.01$ and $0.001$ for the L2 regularisation weight decay and the learning rate, which we then narrowed, testing $0.005$ and $0.0075$ too. For our early stopping we considered values of $10$, $20$ and $30$ for n. After comparing the arithmetic mean of the validation accuracy across the 5 folds we found the hyperparameter combination of $2$ hidden layers with dimensions $(64, 16)$ and an output dimension of $8$ combined with an L2 rate of $0.005$ and a learning rate of $0.01$ with n $= 30$ to have the highest accuracy. Specifically, this had a validation average precision of $81\%$ for predicting correct edges.

In cross validation the early stopping meant the number of epochs was irrelevant (we set it at $300$ just to be an upper ceiling that was never reached due to early stopping). However, when it came to the final training of the model (on the full $80\%$ of non-test set data) we had no validation set to use for early stopping. To overcome this we calculated the average number of epochs in the 5-fold cross validation runs with n $= 30$, which was $82$, and used this number of epochs in our final training. After the final training was complete we found the model had a testing average precision of $78\%$ for predicting correct edges. This being slightly lower than the validation average precision may be surprising given the model is now trained on $80\%$ of the data rather than $64\%$. However, it can be justified as the hyperparameter choices are biased to the validation (training) dataset.

\subsection{Experimental Results}
\begin{figure}[t]
\centering
\includegraphics[scale=0.55]{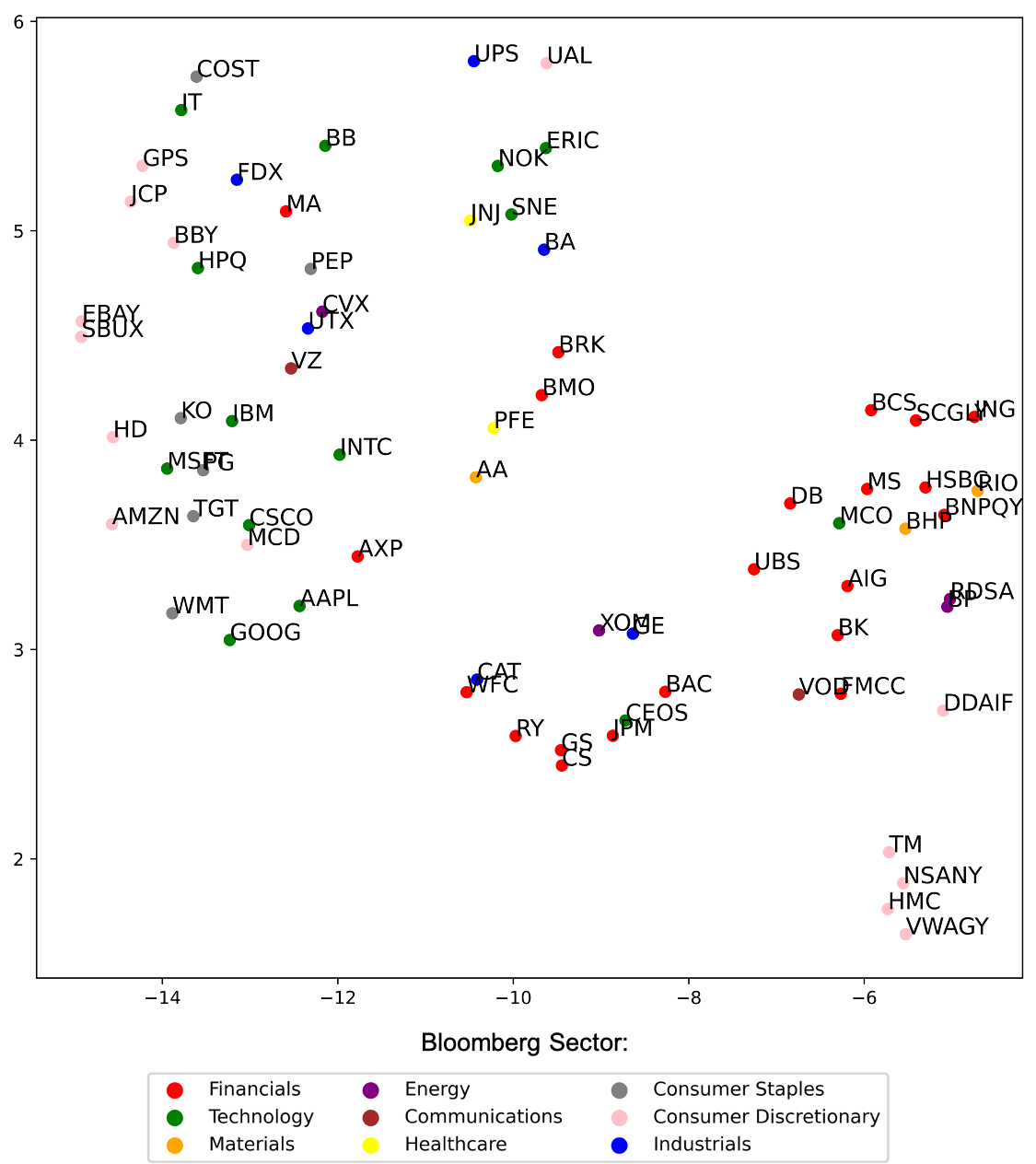}
\caption{\label{fig:company_cluster_plot}Plot showing a $2$ dimension reduction of $\CMcal{G}_\CMcal{F}^{L}$, our finance graph's $8$ dimension latent representation.}
\end{figure}

From our model we took the trained GCN encoder and produced an $8$ dimension latent representation of $\CMcal{G}_\CMcal{F}$, which we notate as $\CMcal{G}_\CMcal{F}^{L}$. Next we wanted to visualise $\CMcal{G}_\CMcal{F}^{L}$ to start to understand the grouping of our companies. To do this we considered various dimension reduction techniques and concluded t-SNE would work best for our data \cite{DBLP:journals/corr/abs-2012-04456}. Figure~\ref{fig:company_cluster_plot} shows a $2$ dimension reduction of $\CMcal{G}_\CMcal{F}^{L}$. Clear clusters of companies from the same Bloomberg sector emerge\footnote[6]{Note the model has no knowledge of the Bloomberg sectors and this partial reconstruction is purely based off of the news co-occurrence and $\mathrm{pvCLCL}$ data.}, such as Volkswagen Group (VWAGY), Nissan (NSANY), Honda Motor (HMC) and Toyota Motor (TM) from the Consumer Discretionary sector. Also clusters that do not match the Bloomberg sectors but are perfectly explainable occur. General Electric (GE, Industrials) and Exxon Mobil (XOM, Energy) provide an example of this, although their Bloomberg sectors differ naturally we would expect them to link as Exxon Mobil are an Oil and Gas company while General Electric produce internal combustion engines.

Given Figure~\ref{fig:company_cluster_plot} seemed reasonable we went back to the $8$ dimension $\CMcal{G}_\CMcal{F}^{L}$ data and applied k-means clustering. We chose k $= 9$ as this matched the number of Bloomberg sectors and produced clusters all containing at least $4$ members. Now that we had our clusters we could beginning comparing them to the Bloomberg sectors as a ground-truth and then to other simpler models to see if we had an increase in performance when considering dual source data. For the simpler models we first consider the GAE trained on just the edge data and then just the vertex data. After this we consider spectral clustering on just the edge data and k-means clustering on just the vertex data.

\begin{figure}[b]
\centering
\includegraphics[scale=0.6]{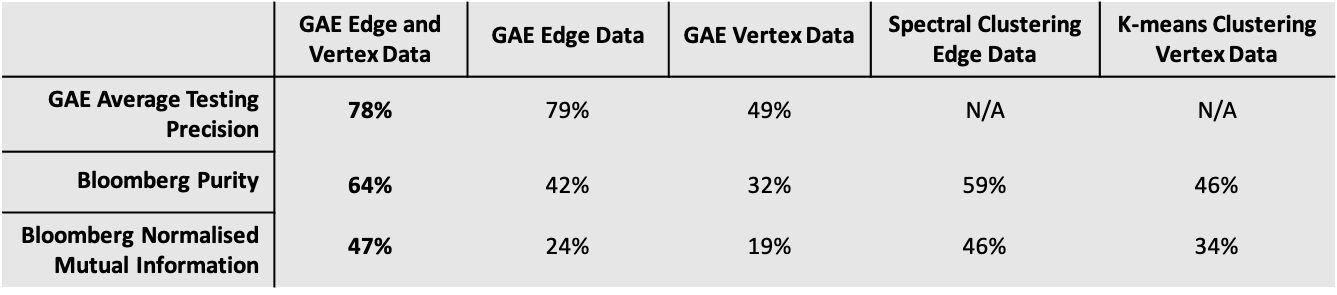}
\caption{\label{fig:quantitative_comparison_results}Table showing the results of the cluster produced from our GAE trained on $\CMcal{G}_\CMcal{F}$ (i.e. edge and vertex data) alongside the clusters produced from simpler methods.}
\end{figure}

Figure~\ref{fig:quantitative_comparison_results} shows the quantitative results of our cluster comparisons with the other methods. Firstly, we compared our clusters produced from the GAE with the full finance graph (i.e. edge and vertex data) to the Bloomberg sectors. This led to a Purity score of $64\%$ and a Normalised Mutual Information score of $47\%$. These scores are both reasonably high and show our clusters are well informed. These clusters could now be used for downstream finance tasks that specifically require companies to be grouped by news co-occurrence, market close data and the interactions between the two.

Next we compared our results to the same GAE model but trained on just the edge data and then on just the vertex features. To train and run the GAE on just the edge data ($\CMcal{G}_\CMcal{F}$ without any vertex features) we use $\mathbf{I}_{72}$ in place of $\mathbf{X_\CMcal{F}}$ as some form of vertex features are needed for the model to run. This obtains an average testing precision of $79\%$ which is very similar to our model trained on $\CMcal{G}_\CMcal{F}$. However, when we apply k-means clustering with k $= 9$ to the latent representation we find it only has a purity score of $42\%$ and a Normalised Mutual Information score of $24\%$ when compared to the Bloomberg sectors. This discrepancy between average testing precision and the cluster metrics can be explained due to the fact testing precision is purely based off how well the GAE can reconstruct the graph given, not whether useful information has actually been produced.

To train and run the GAE on just the vertex features we take $\CMcal{G}_\CMcal{F}$ and then randomise the edges (taking care to not produce any self-loops or assign two edges to the same vertex pair). Once trained on this randomised edge graph the model obtained an average testing precision of $49\%$. After applying k-means clustering with k $= 9$ to the latent representation the clustering had a purity score of $32\%$ and a Normalised Mutual Information score of $19\%$ when compared to the Bloomberg sectors. All these metrics are significantly lower and thus show the importance of the news co-occurrence edges to the model.

Now we wanted to see how our results compared to simpler more traditional methods of clustering data. These methods cannot incorporate both vertex features and edge weights thus we use two comparisons, one for each of the datasets. Firstly, we take just the edge weights, that is we consider $\CMcal{G}_\CMcal{F}$ without any vertex features added. To this graph we apply spectral clustering to produce 9 clusters. Comparing this clustering with the Bloomberg sectors led to a Purity score of $59\%$ and a Normalised Mutual Information score of $46\%$.

Secondly, we consider just the vertex features where we apply k-means clustering to the vertex feature vectors with k $= 9$. Comparing this clustering with the Bloomberg sectors led to a Purity score of $46\%$ and a Normalised Mutual Information score of $34\%$. Thus, both the simple traditional methods produce worse results than our clustering when compared to the ground-truth Bloomberg sectors. Spectral clustering still performs relatively well; this makes sense as not only is spectral clustering a hugely successful technique but the majority of our data's relational information comes from just the edge weights.

\begin{figure}[t]
\centering
\includegraphics[scale=0.45]{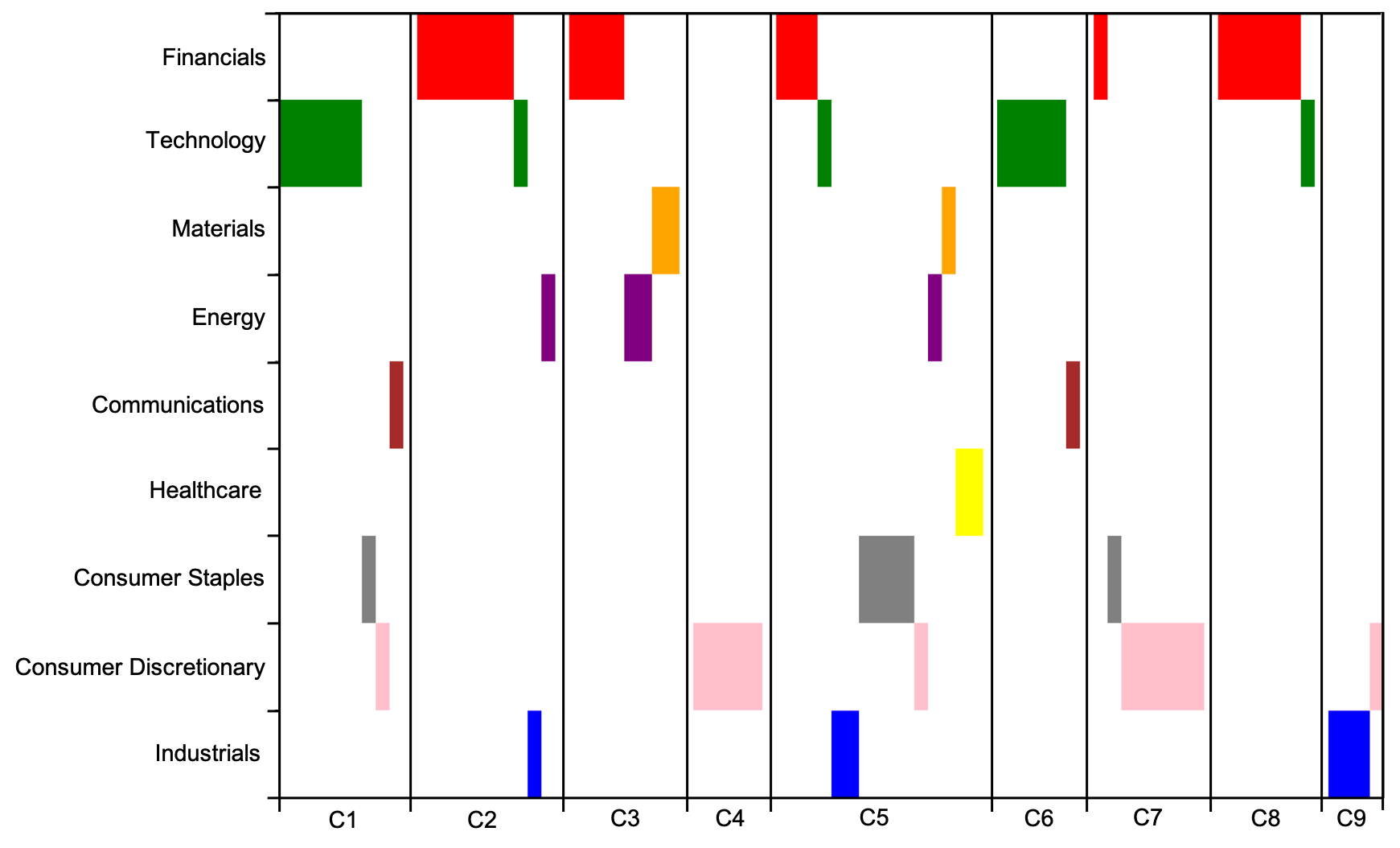}
\caption{\label{fig:cluster_comparison_plot}Plot showing the 9 sections (C1, ..., C9) of our company clustering decomposed into the Bloomberg Sectors. For each section of our cluster the companies within it are each represented by a rectangle with vertical position and colour matching the associated Bloomberg sector.}
\end{figure}

Figure~\ref{fig:cluster_comparison_plot} shows the cluster produced from our GAE trained on $\CMcal{G}_\CMcal{F}$ decomposed into the Bloomberg sectors. This visualisation shows how some of our cluster sections compose of just one Bloomberg sector (C4 contains just Consumer Discretionary companies) while other have grouped companies across multiple sectors.

\section{Conclusion and Future Work}
We have shown that GAEs are able to extract both edge information (in the form of news co-occurrence data) and vertex features (in the form of optimised market close data). We then showed how the trained encoder can be used to produce a sensible latent representation of the graph. We also illustrated how after clustering the latent representation we can produce clusters comparable to that of industry standards but which have been formed from specific chosen data. We demonstrated that such models are able to perform with high accuracy on relatively small datasets and outperform both classical graph clustering models and also GAEs with only one data source. The framework we provide outlines how to appropriately clean data and implement and train a GAE for future work. 

We feel an interesting next step would be to extend the vertex features. $250$ dimensions is relatively low for such a setup thus a natural extension is to include the $2008$ (and possibly further) market close data\footnote[7]{Here one would likely want to include news articles from the additional years when calculating the news co-occurrence matrix too. We would take caution actually using $2008$ data as this was a period of unpredictability and high volatility, instead we would suggest starting again at a new later period.}. Also, we do not take into account the time-series nature of our vertex features due to this not being overly relevant to the task of clustering and requiring more complex models. That said, for future work we believe explicitly taking into account the time-series aspect would be an appropriate next step. This would provide a solid foundation for a vertex prediction model (i.e. predicting company's future stock price). A framework such as EvolveGCN or a Spatio-Temporal Graph Convolutional Network would allow for such a task \cite{DBLP:journals/corr/abs-1902-10191, DBLP:journals/corr/abs-1709-04875}.

\newpage
\section*{Acknowledgements}
\label{sec:Acknowledgments}
The author would like to thank the Oxford-Man Institute for their generous funding of this project. He would also like to thank Prof. Mihai Cucuringu and Prof. Xiaowen Dong for taking the time to supervise this project and their kind support throughout. The author also gives thanks Xingchen Wan for providing and allowing use of their data.

\section*{Data Availability}
Data, materials and code that are necessary to reproduce the results described in this paper are available from the author upon reasonable request. A large proportion of the code used is also available on \href{https://github.com/edwardturner01/graph_auto-encoders_for_financial_clustering}{github.com/edwardturner01/graph\_auto-encoders\_for\_financial\_clustering} and market data of equivalent quality used in this study are publicly available.

\bibliographystyle{unsrt}  
\bibliography{references}

\begin{thebibliography}{10}

\bibitem{app9245574}
Francesco Rundo, Francesca Trenta, Agatino~Luigi di~Stallo, and Sebastiano
  Battiato.
\newblock Machine learning for quantitative finance applications: A survey.
\newblock {\em Applied Sciences}, 9(24), 2019.

\bibitem{grl}
William~L. Hamilton.
\newblock Graph representation learning.
\newblock {\em Synthesis Lectures on Artificial Intelligence and Machine
  Learning}, 14(3):1--159, 2017.

\bibitem{Goodfellow-et-al-2016}
Ian Goodfellow, Yoshua Bengio, and Aaron Courville.
\newblock {\em Deep Learning}.
\newblock MIT Press, 2016.
\newblock \url{http://www.deeplearningbook.org}.

\bibitem{DBLP:journals/corr/KipfW16}
Thomas~N. Kipf and Max Welling.
\newblock Semi-supervised classification with graph convolutional networks.
\newblock {\em CoRR}, abs/1609.02907, 2016.

\bibitem{kipf2016variational}
Thomas~N. Kipf and Max Welling.
\newblock Variational graph auto-encoders, 2016.

\bibitem{DBLP:journals/corr/abs-1909-10660}
Daiki Matsunaga, Toyotaro Suzumura, and Toshihiro Takahashi.
\newblock Exploring graph neural networks for stock market predictions with
  rolling window analysis.
\newblock {\em CoRR}, abs/1909.10660, 2019.

\bibitem{DBLP:journals/corr/abs-2011-06430}
Xingchen Wan, Jie Yang, Slavi Marinov, Jan{-}Peter Calliess, Stefan Zohren, and
  Xiaowen Dong.
\newblock Sentiment diffusion in financial news networks and associated market
  movements.
\newblock {\em CoRR}, abs/2011.06430, 2020.

\bibitem{kingma2017adam}
Diederik~P. Kingma and Jimmy Ba.
\newblock Adam: A method for stochastic optimization, 2017.

\bibitem{DBLP:journals/corr/abs-1205-2653}
Corinna Cortes, Mehryar Mohri, and Afshin Rostamizadeh.
\newblock {L2} regularization for learning kernels.
\newblock {\em CoRR}, abs/1205.2653, 2012.

\bibitem{DBLP:journals/corr/abs-2012-04456}
Yingfan Wang, Haiyang Huang, Cynthia Rudin, and Yaron Shaposhnik.
\newblock Understanding how dimension reduction tools work: An empirical
  approach to deciphering t-sne, umap, trimap, and pacmap for data
  visualization.
\newblock {\em CoRR}, abs/2012.04456, 2020.

\bibitem{DBLP:journals/corr/abs-1902-10191}
Aldo Pareja, Giacomo Domeniconi, Jie Chen, Tengfei Ma, Toyotaro Suzumura,
  Hiroki Kanezashi, Tim Kaler, and Charles~E. Leiserson.
\newblock Evolvegcn: Evolving graph convolutional networks for dynamic graphs.
\newblock {\em CoRR}, abs/1902.10191, 2019.

\bibitem{DBLP:journals/corr/abs-1709-04875}
Bing Yu, Haoteng Yin, and Zhanxing Zhu.
\newblock Spatio-temporal graph convolutional neural network: {A} deep learning
  framework for traffic forecasting.
\newblock {\em CoRR}, abs/1709.04875, 2017.

\end{thebibliography}

\newpage

\centerline{\large \bfseries \scshape Appendix}

\section*{A.1 \space Unwindorised Node Feature Heatmaps}
\label{sec:Appendix}

\begin{figure}[h!]
\centering
\includegraphics[scale=4.2]{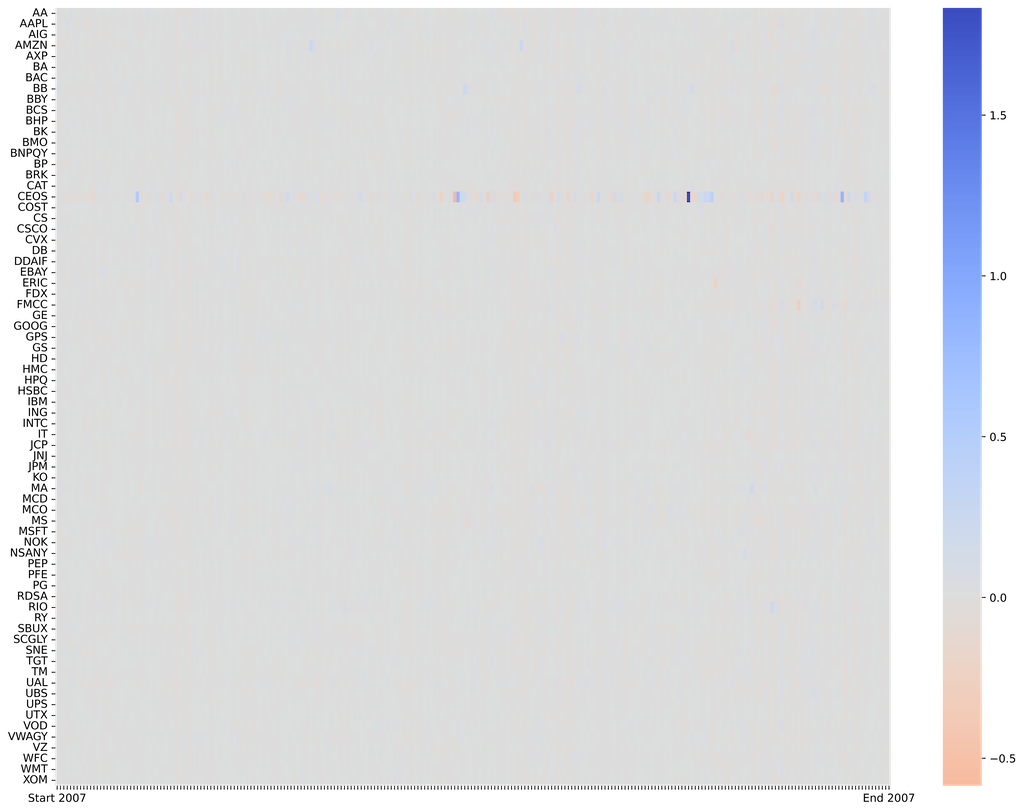}
\caption{\label{fig:original_heatmap}Heatmap of $\mathrm{pvCLCL}$ vertex features. Due to the extreme anomalies from CeCors the majority of entries appear grey as the scale is distorted.}
\end{figure}

\begin{figure}[h!]
\centering
\includegraphics[scale=4.2]{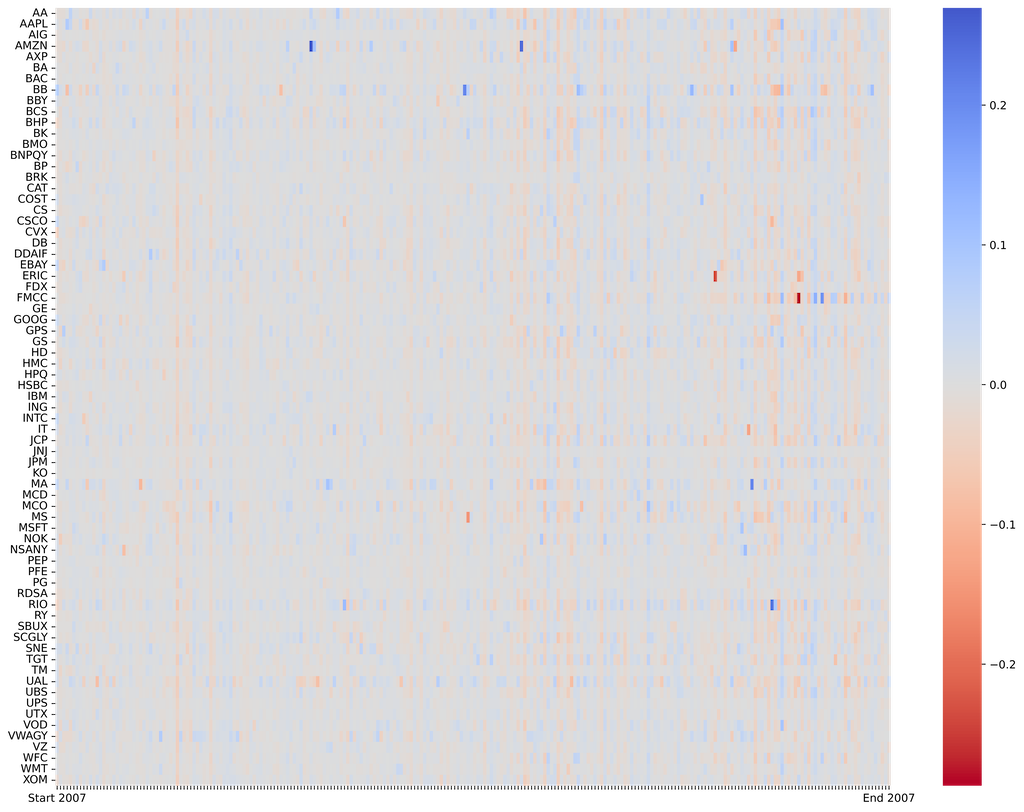}
\caption{\label{fig:without_CEOS_heatmap}Heatmap of $\mathrm{pvCLCL}$ vertex features with CeCors (CEOS) removed. Now the scaling is less distorted and a greater mix of colours can be seen.}
\end{figure}

\section*{A.2 \space List of Companies Used}

\begin{center}
\begin{longtable}{|l|l|l|}
\caption{The Bloomberg sectors of the 72 companies used and their tickers\label{tab:Companies_tickers}}\\

\hline \multicolumn{1}{|c|}{\textbf{Bloomberg Sector}} & \multicolumn{1}{c|}{\textbf{Ticker}} & \multicolumn{1}{c|}{\textbf{Company}} \\ \hline 
\endfirsthead
\multicolumn{3}{c}
{{ \tablename\ \thetable{} -- Continued from previous page}} \\
\hline \multicolumn{1}{|c|}{\textbf{Bloomberg Sector}} & \multicolumn{1}{c|}{\textbf{Ticker}} & \multicolumn{1}{c|}{\textbf{Company}} \\ \hline 
\endhead
\endfoot
\hline 
\endlastfoot
\hline

\multirow{21}*{Financials (21)}&AXP&American Express\\
&AIG&American International Group\\
&BAC&Bank of America\\
&BMO&Bank of Montreal\\
&BK&Bank of New York Mellon\\
&BCS&Barclays\\
&BRK&Berkshire Hathaway\\
&BNPQY&BNP\\
&CS&Credit Suisse\\
&DB&Deutsche Bank\\
&FMCC&Federal Home Loan Mortgage Corp\\
&GS&Goldman Sachs\\
&HSBC&HSBC\\
&ING&ING Group\\
&JPM&JP Morgan\\
&MA&MasterCard\\
&MS&Morgan Stanley\\
&RY&Royal Bank of Canada\\
&SCGLY&Societe Generale\\
&UBS&UBS\\
&WFC&Wells Fargo\\
\hline

\multirow{14}*{Technology (14)}
&AAPL&Apple\\
&BB&BlackBerry\\
&CEOS&CeCors\\
&CSCO&Cisco\\
&ERIC&Ericsson\\
&IT&Gartner\\
&GOOG&Google\\
&HPQ&Hewlett Packard\\
&IBM&IBM\\
&INTC&Intel\\
&MSFT&Microsoft\\
&MCO&Moody's Corporation\\
&NOK&Nokia\\
&SNE&Sony\\
\hline

\multirow{3}*{Materials (3)}
&AA&Alcoa\\
&BHP&BHP\\
&RIO&Rio Tinto\\
\hline

\multirow{4}*{Energy (4)}
&BP&BP\\
&CVX&Chevron\\
&RDSA&Royal Dutch Shell\\
&XOM&Exxon Mobil\\
\hline

\multirow{2}*{Communications (2)}
&VZ&Verizon Communications\\
&VOD&Vodafone\\
\hline

\multirow{2}*{Healthcare (2)}
&JNJ&Johnson \& Johnson\\
&PFE&Pfizer\\
\hline

\multirow{6}*{Consumer Staples (6)}
&KO&Coca-Cola\\
&COST&Costco Wholesale\\
&PEP&Pepsi\\
&PG&Procter \& Gamble\\
&TGT&Target Corporation\\
&WMT&Walmart\\
\hline

\newpage

\multirow{14}*{Consumer Discretionary (14)}
&AMZN&Amazon\\
&BBY&Best Buy\\
&DDAIF&Daimler AG\\
&EBAY&Ebay\\
&GPS&Gap Inc.\\
&HD&Home Depot\\
&HMC&Honda Motor\\
&JCP&JC Penney\\
&MCD&McDonald\\
&NSANY&Nissan\\
&SBUX&Starbucks\\
&TM&Toyota Motor\\
&UAL&United Airlines Holdings\\
&VWAGY&Volkswagen Group\\
\hline

\multirow{6}*{Industrials (6)}
&BA&Boeing\\
&CAT&Caterpillar Inc.\\
&FDX&FedEx\\
&GE&General Electric\\
&UPS&United Parcel Service\\
&UTX&United Technologies\\

\end{longtable}
\end{center}

\end{document}